# Neutral particle analyzer for plasma diagnostics on tokamak ST40

**S. Polosatkin,**[a,b,c*] **V. Belykh**[a], **A. Rovenskikh**[a]

[a] *Budker Institute of Nuclear Physics,*
   *630090, Lavrent'eva av.11, Novosibirsk, Russia*

[b] *Novosibirsk State Technical University,*
   *630073, Prospekt K. Marksa 20, Novosibirsk, Russia*

[c] *Novosibirsk State University,*
   *630090, Pirogova str. 1, Novosibirsk, Russia*

   *E-mail:* s.v.polosatkin@inp.nsk.su

ABSTRACT: The paper presents a detailed description of a neutral particle analyzer designed and produced for plasma diagnostics on the spherical tokamak ST40. The aim of the diagnostic is to measure both the bulk ion temperature in the range from 0.5 to 10 keV and the energy distribution of fast ions with energies up to 40 keV, which appear in the plasma through neutral beam injection. A feature of the analyzer is its ability to separate hydrogen isotopes (protium and deuterium) and measure the ion distribution function of the selected isotope.

KEYWORDS: Nuclear fusion; Plasma diagnostics; Neutral particle analyzer.

---

[*] Corresponding author.

## Introduction

Analysis of energy distribution of charge-exchange neutrals is an informative tool for fusion plasma diagnostics. Both passive and active (beam-assisted) diagnostics of charge-exchange particles are widely used on most plasma facilities; different types of neutral particle analyzers (NPAs), which use electric or magnetic fields for particle separation, have been developed [1-3]. The paper presents a detailed description of an NPA designed and manufactured for plasma diagnostics on the spherical tokamak ST40 [4]. The goal of this diagnostic is to measure both the bulk ion temperature in the range of 0.5–10 keV and the energy distribution of fast ions that appear in the plasma due to neutral beam injection. The target parameters of the designed NPA are as follows:

- Possibility of measuring the energy distribution function of either hydrogen or deuterium neutrals leaving the plasma.
- Easy switching between the detection of hydrogen and deuterium without disassembling the NPA.
- Energy range: 0–40 keV for hydrogen and 0–20 keV for deuterium neutrals.
- Energy resolution: 10%.
- Temporal resolution: 1 ms.
- Minimum measurable temperature: 500 eV.
- Potential for future upgrades to increase the energy range for hydrogen neutrals up to 70 keV and enable simultaneous detection of both hydrogen and deuterium neutrals.

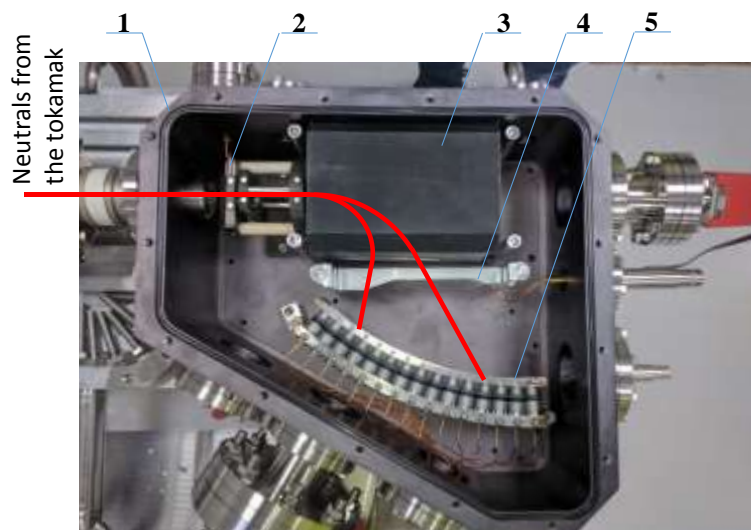

**Figure 1.** The NPA interior; 1 – the NPA body, 2 – stripping foil unit, 3 – bending magnet, 4 – deflector plates, 5 – detector unit

## 1. The general layout of the NPA

The general layout of the NPA replicates the schemes of analyzers previously designed for MST and C-2 facilities [5] (see Fig. 1). Neutrals leaving the plasma are ionized in the stripping foil, accelerated by the voltage applied to the foil, and, after consecutively flying through a bending magnet and electrostatic deflector plates, are counted by a row of detectors vertically shifted relative to the central plane. A soft iron vacuum chamber is used to shield the NPA's interior from the stray magnetic field of the tokamak. The analyzer is equipped with a Bayard-Alpert-type ion source for on-site calibration of the NPA and checking the foil's condition.

## 1.1 Stripping foil unit

An ultrathin (10 nm) solid carbon film is used in the NPA for converting (stripping) neutrals leaving the plasma into ions, which can then be separated in magnetic and electric fields. Using a solid target instead of a gaseous one reduces the gas load and simplifies the design and operation of the NPA. Specifically, applying a positive voltage to the film allows adjusting the energy range of the analyzer and matching the energy dispersion with the geometric dimensions of the detectors. However, a drawback of this scheme is the excessive scattering of particles in the foil (particularly significant for low-energy neutrals), which affects the detection efficiency and NPA resolution.

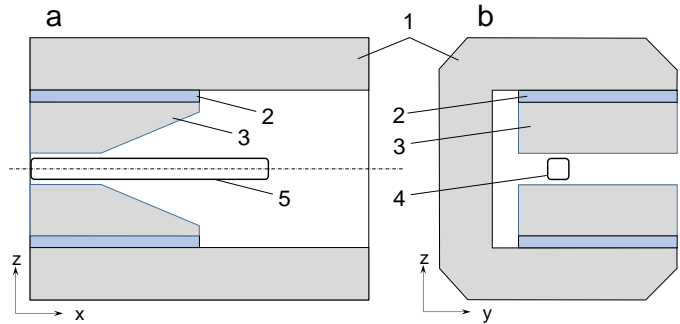

**Figure 2.** The bending magnet. a – side view, b - front view; 1 – magnet yoke, 2 – permanent magnets, 3 – magnet pole, 4 – input aperture, 5 – output aperture (bodies input and output diaphragms are not shown)

## 1.2 Bending magnet and deflector plates

A drawing of the bending magnet is shown in Fig. 2. Two sets of 4.5-mm-wide permanent NdFeB magnets (N38SH grade) serve as sources of magnetic flux. The magnet poles create a magnetic field that decreases along the x-axis (the direction of the primary neutrals' velocity). The magnetic field distribution along the central axis of the NPA is depicted in Fig. 3. This magnetic field profile enhances the angular dispersion of ions in the high-energy portion of the operational energy range, providing the necessary energy resolution. The position of ions on the detector surface versus their energy is presented in Fig. 3b. The linear dispersion for hydrogen ions ranges from 0.15 (at lower energies) to 0.35 (at higher energies) keV/mm.

One notable feature of the NPA is its capability to differentiate between specific hydrogen isotopes being measured. This functionality is enabled by using parallel magnetic and electric fields for energy and mass dispersion (Thomson parabola scheme [6]). The specifically calculated shape of the deflector plates ensures a consistent vertical displacement for ions with a given mass but varying energies on the detector surface.

It is straightforward to deduce that the trajectory of an ion within the bending magnet depends on the product of its mass and energy.

$$\vec{r} = f\left(\frac{mE}{q^2}\right)$$

while deflection in the electrostatic field in the vertical (xz) plane is inversely proportional to the energy of the ion

$$\varphi \sim \frac{q}{E}.$$

Here, $E$ represents the energy of the ion in the bending magnet (note that the energy of the ion differs from the energy of the primary neutral by the foil voltage and energy loss in the foil), while $m$ and $q$ – mass, and charge of the ion. So, the energies of hydrogen and deuterium ions moving along the same trajectory in the bending magnet differ by a factor of two, and the vertical tilt angle in the deflector for a given trajectory ($\frac{mE}{q^2} = const$) is proportional to the mass of the isotope. Therefore, the selection of the measuring isotope is achieved by matching changes in the deflector plate voltage (-8 kV for H and -4 kV for D) and the foil voltage (+8 kV for H and +4 kV for D).

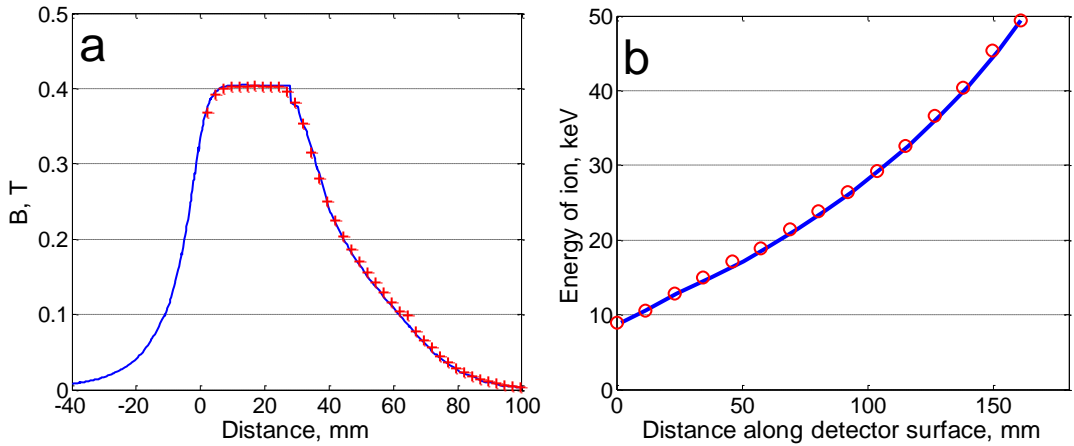

**Figure 3. a** – magnetic field along the central axis of NPA, solid line – calculation, crosses – measurements; Bending magnet. a – side view, b - front view; 1 – magnet yoke, 2 – permanent magnets, 3 – magnet pole, 4 – input aperture, 5 – output aperture (input and output diaphragms are not shown); **b** – coordinates of hydrogen ions with different energies on the detector surface, solid line – calculation, rings - measurements

### 1.3 Detector unit

A channeltron electron multiplier MAGNUM 5900 from Photonis [7] was chosen as a tool for particle detection because of its high output current, which allows increasing the dynamic range of the NPA. Fifteen channeltrons used for particle detection are placed in a radial array to provide the closest spacing of the detectors. Detectors' entrance apertures are arranged on a cylindrical surface with a radius of 220 mm, and the distance between detectors is 11.5 mm. Since low-energy ions acquire a large angular spread in the stripping foil, the low-energy side of the detector row should be placed closer to the bending magnet, while the high-energy side should be farther from the magnet to increase linear dispersion.

### 1.4 Calibration ion source

Interpretation of measurements with the NPA and determination of ion temperature requires information about the instrumental functions of the NPA channels. The NPA is equipped with a special low-current electron ionization ion source for measuring the instrumental functions and controlling the stripping foil's consistency. The overall design of the ion source is similar to the construction of a hot cathode (Bayard-Alpert) vacuum gauge. A unique feature of the design is that the ion source is mounted directly on the axis of the beamline and is made transparent for fast neutrals coming from the plasma.

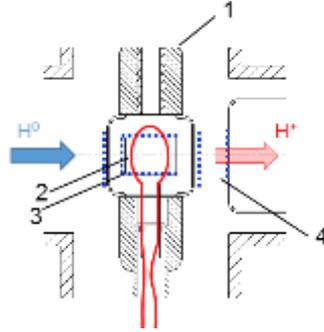

**Figure 4.** Calibration ion source. 1 – high voltage terminal, 2 – hot filament, 3 – ion source grid, 4 – accelerating gap

A scheme of the ion source is presented in the Fig. 4. The ion source consists of a hot cathode filament and a cylindrical grid mounted inside the ion source housing. The cathode and the grid are biased, accordingly, at +50 and +200 V relative to the housing. Electrons emitted from the cathode are accelerated and oscillate in the electrostatic potential well formed by the grid, producing ions. These ions are extracted and accelerated by a high voltage (up to -25 kV) applied to the housing. On-axis face parts of the ion source are made of fine mesh to provide transparency for neutrals while maintaining uniformity of the electric field in the accelerating gap. The operational pressure range of the ion source is $10^{-5} - 10^{-3}$ Pa. Although the flux of H$^+$ ions produced from residual gas at a pressure of $10^{-5}$ Pa is sufficient for calibrating the NPA, additional gas can be supplied directly inside the ion source housing if needed.

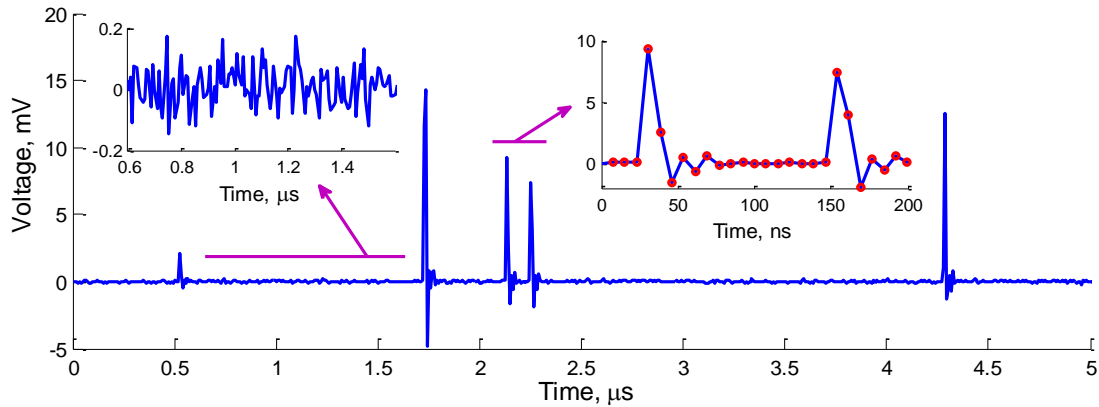

**Figure 5.** Part of waveform of the NPA detection channel in the experiment on ST40 (the signals was inverted for convenience). Channeltron voltage 2 kV, full span of the digitizer 400 mV. Noise level and a shape of pulse have shown on the insets.

## 2. Particle flux registration

MAGNUM 5900 channeltron electron multipliers have been used in the NPA for particle detection, allowing operation in both current and pulse-counting modes. Pulse counting is chosen as the main operation mode of the NPA due to its high sensitivity and the absence of problems with channel nonlinearity and differences in channeltron gains. Since the pulse duration of the ST40 does not exceed several seconds, we decided to apply direct fast signal digitization with offline post-processing for pulse counting. The SPECTRUM DN6.441-16 16-channel 130 MHz digitizing system with 16-bit resolution is used for signal recording. Note that using the digitizer leaves open the possibility of operating in current mode with a temporal resolution of about a few hundred microseconds, as well as implementing pulse-counting algorithms in the built-in FPGA of the digitizer.

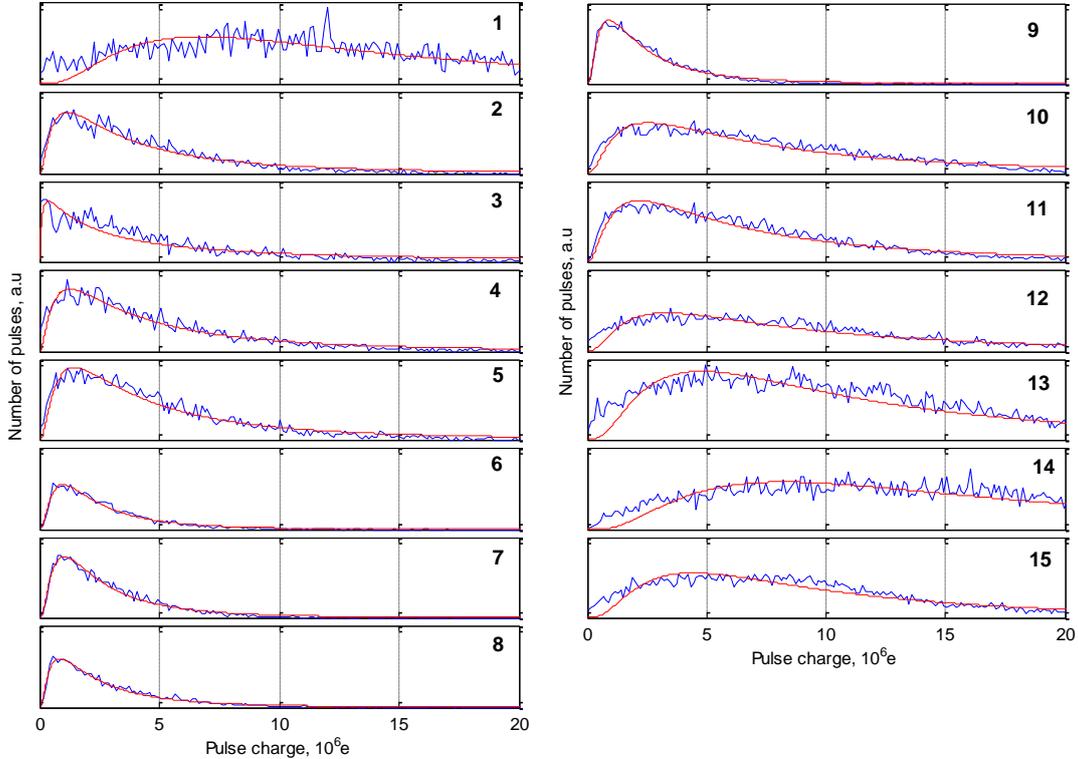

**Figure 6**. Pulse height distributions for the NPA channels (ST40 shot #6141). Red curves – approximations by the lognormal function.

A part of the typical waveform from one of the detection channels of the NPA, recorded during a standard ST40 shot, is shown in Fig. 5. The RMS noise of the baseline of the signal is 70 µV, while the peak amplitude of the pulses reaches several mV, and their duration is about ten nanoseconds. Pulse height distributions for the NPA channels are presented in Fig. 6. These observed pulse height distributions can be reasonably approximated by a lognormal function, as indicated by the red curves in Fig. 6.

### 3. Instrumental functions evaluation

### 3.1 Measurements of instrumental functions of low-energy channels

The result of calibration for the first eight channels of the NPA is shown in Fig. 7. The ions were generated from the residual vacuum, with an operating pressure of $10^{-5}$ Pa, and the foil voltage and deflector voltage were set to $U_{foil}$=+8 kV and $U_{def}$=-8 kV, respectively. The two main ion species produced by the ion source are $H^+$ and $H_2^+$. The latter completely dissociates in the stripping foil, producing half-energy protons. Different species require different initial energies to hit the same point on the detector surface. Fast neutrals from the plasma reach the stripping foil without any change in energy, lose some energy ($\Delta E_{loss}$) in the foil, get ionized, and then accelerate due to the foil bias voltage ($U_{foil}$). In contrast, $H^+$ ions generated by the ion source first decelerate in the gap between the NPA entrance and the foil, and then re-accelerate after passing through the

– 5 –

foil. For $H_2^+$ ions, the energy gained in the electric field before dissociation is shared equally between the two protons forming the molecule. After passing through the foil and undergoing dissociation, both protons are accelerated to the full foil voltage. Therefore, there exists the following relationship between the initial energies of different types of particles hitting the same spot on the detector surface:

$E_{H0} = E_{H+} - eU_{foil} = 0.5*(E_{H2+} - eU_{foil})$

Two peaks corresponding to $H^+$ and $H_2^+$ ions overlap in the first NPA channel, where the equivalent neutral energy is close to the foil voltage, and they separate in subsequent higher-energy channels. The intensities of the $H^+$ and $H_2^+$ peaks are comparable when the ion source operates with the residual vacuum, unlike the case of measurements with additional hydrogen puffing, where the $H_2^+$ peak exceeds the $H^+$ peak by more than tenfold. The background in channel 1 is attributed to $H_2O^+$ and other molecular ions that dissociate in the foil, whereas the background in channel 4 is linked to low-energy $H^-$ ions that transform into protons in the foil and thus experience double acceleration by the foil voltage.

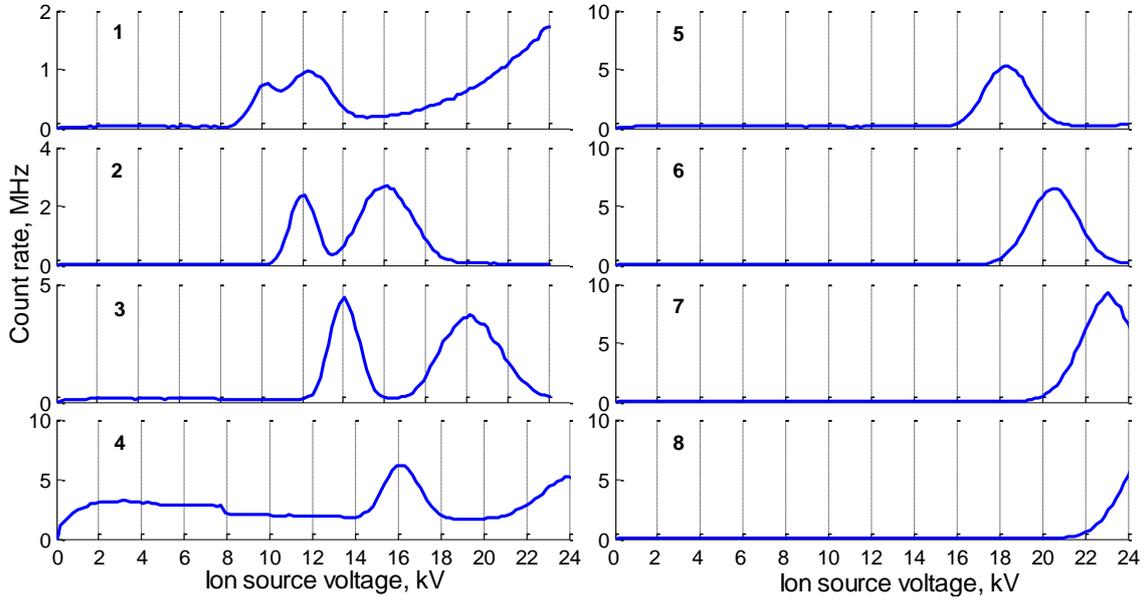

**Figure 7.** Response of NPA channels 1-8 for different accelerating voltage of calibration ion source. Stripping foil bias voltage +8 kV, deflection voltage -8 kV, ion source operated with residual vacuum $10^{-5}$ Pa

The instrumental functions are well approximated by Gaussian curves. The mean equivalent energies of neutrals, FWHM widths, and relative amplitudes of the measured instrumental functions for NPA channels 1-8 are presented in Table 1.:

Table 1:

| Channel | Measurements $H^0$ | | | Calculations $H^0$ | | | Calculations $D^0$ | | |
|---|---|---|---|---|---|---|---|---|---|
| | Equivalent energy of $H_0$, keV | FWHM width, keV | Peak efficiency | Equivalent energy of $H_0$, keV | FWHM width, keV | Peak efficiency | Equivalent energy of $D_0$, keV | FWHM width, keV | Peak efficiency |

| 1  | 2    | 1.4 | 2.5e-3 | 1.93 | 1.1  | 1.9e-3 | 1.1  | 0.85 | 3.6e-4 |
|----|------|-----|--------|------|------|--------|------|------|--------|
| 2  | 4    | 1.4 | 9.8e-3 | 4    | 1.1  | 6e-3   | 2    | 0.79 | 7.6e-4 |
| 3  | 6    | 1.7 | 1.8e-2 | 6.06 | 1.4  | 1.1e-2 | 3.1  | 1    | 1.3e-3 |
| 4  | 8.1  | 1.7 | 1.8e-2 | 8.13 | 1.66 | 1.8e-2 | 4.2  | 1.1  | 2e-3   |
| 5  | 10.3 | 2.2 | 2.1e-2 | 10.2 | 1.9  | 2.6e-2 | 5.2  | 1.2  | 3.6e-3 |
| 6  | 12.6 | 2.5 | 2.6e-2 | 12.6 | 2.2  | 3.1e-2 | 6.4  | 1.6  | 3.6e-3 |
| 7  | 15   | 2.7 | 3.8e-2 | 14.9 | 2.4  | 3.8e-2 | 7.6  | 1.7  | 4.3e-3 |
| 8  | 17.5 | 3.2 | 4.1e-2 | 17.6 | 2.6  | 4.2e-2 | 8.9  | 2.2  | 4.7e-3 |
| 9  |      |     |        | 20.6 | 2.9  | 4.5e-2 | 10.3 | 2.2  | 5.2e-3 |
| 10 |      |     |        | 23.7 | 3.1  | 5.3e-2 | 12.1 | 2.5  | 7.4e-3 |
| 11 |      |     |        | 27.1 | 3.4  | 6.1e-2 | 13.7 | 2.9  | 8.8e-3 |
| 12 |      |     |        | 30.7 | 3.6  | 6.5e-2 | 15.6 | 3.1  | 1.1e-2 |
| 13 |      |     |        | 35.1 | 3.7  | 7.1e-2 | 18   | 3.4  | 1.1e-2 |
| 14 |      |     |        | 39.8 | 4.2  | 7.5e-2 | 20.3 | 3.8  | 1.4e-2 |
| 15 |      |     |        | 44.9 | 4.4  | 8.9e-2 | 23.1 | 4.1  | 1.5e-2 |

**3.2 Stripping foil thickness evaluation**

Maximum operational voltage of the calibration source is insufficient for direct calibration of high-energy channels using hydrogen ions. These channels were intended to be calibrated using deuterium ions, but operation with deuterium was not allowed due to safety concerns during NPA calibration. Therefore, calculations of particle trajectories within the NPA were utilized to evaluate the instrumental functions of the high-energy channels.

The calculations are conducted in three stages:
- Calculation of transmission and scattering of ions in the stripping foil
- Calculation of particle trajectories in the magnetic and electric fields of the analyzer
- Post-processing of trajectory data and determining the probability of counting incoming particles with specific energy by NPA channels

During the first stage, the passage of ions with a given initial energy through the stripping foil is computed using the Monte Carlo code SRIM [8]. The output data from the SRIM calculation—energies and scattering angles of test particles transmitted through the foil—are employed as input parameters for subsequent simulation stages.

An essential aspect of these calculations is ensuring the accuracy of simulating ion interactions with foil material. We employ a 10-nm-thick (2 μg/cm²) carbon foil supported by a nickel mesh from Lebow Company [9] for ion stripping. Although the foil's thickness is monitored and assured by suppliers with a precision of 10%, it can fluctuate during storage and handling. Additionally, SRIM calculations involve multiple parameters of ion-material interactions that might differ for ultra-thin foils compared to bulk carbon. Hence, we established the "effective" foil thickness by comparing the energy loss in the foil, as observed experimentally during NPA calibration, with calculations for various foil thicknesses.

Energy losses were inferred from measurements of the NPA detectors' responses during scans of the accelerating voltage on the calibration ion source. The results of such calibration for NPA Channel 3 are depicted in Figure 8. Curve 1 in this figure was obtained when a partially damaged foil with through-holes was inserted into the NPA. During this run, the foil bias voltage was set to zero, meaning the proton energy upon entering the foil equated to the ion accelerating voltage. Two distinct peaks evident on the curve represent ions that traversed the foil's holes (with no energy loss) and those transmitted through the foil, losing part of their initial energy. The observed energy loss of approximately 1.6 keV notably surpasses expectations based on the stopping power value for 14 keV protons (0.46 keV/(g/cm²)) [10] and the nominal foil linear density of 2 μg/cm², corresponding to an effective foil thickness of 17 nm.

There is a likelihood that foil damage increased the thickness of intact sections. To validate this hypothesis, we performed supplementary measurements using a fresh, undamaged foil (Curve 2 in Figure 8). During this run, the foil bias voltage was +8 kV, implying that for an ion source accelerating voltage of 14 kV, the proton energy upon entering the foil would be 6 keV. The observed energy loss in the foil for this run amounted to 1.1 kV, and the stopping power for 6 keV protons is 0.33 keV/(g/cm²), yielding the same effective foil thickness of 17 nm.

Maximal operation voltage of the calibration source is insufficient for direct calibration of high-energy channels using hydrogen ions. These channels were supposed to be calibrated using deuterium ions, but operation with deuterium was not permitted because of safety issues when NPA calibration was produced. Therefore calculation of particles trajectories in the NPA was used for the evaluation of instrumental functions of high-energy channels.
The calculations are carried out in three stages:
- calculation of transmission and scattering of ions in the stripping foil
- calculation of particles trajectories in the magnetic and electric fields of the analyzer
- postprocessing of trajectories data and determination of the probability of counting of the incoming particles with certain energy by NPA channels

In the first stage passage of ions with certain initial energy via stripping foil is calculated by Monte-Carlo code SRIM [8]. Output data of SRIM calculation – energies and scattering angles of test particles transmitted through the foil - are used as input parameters for the next stages of the simulation.

An important issue of such calculations is the correctness of the simulation of ion interaction with foil matter. We use 10 nm (2 μg/cm$^2$) carbon foil on the nickel support mesh from Lebow company [9] for ion stripping. Even though foil thickness is controlled and guaranteed by supplies on the accuracy level of 10% the foil thickness can vary during storage and handling. Moreover, SRIM calculations use several parameters of ion-mater interaction which may differ for ultrathin foil and bulk carbon. Therefore, we determined the "effective" foil thickness by comparison energy loss in the foil experimentally observed in NPA calibration with calculations for different foil thicknesses.

The energy loss was derived from measurements of the response of the NPA detectors during a scan of accelerating voltage on the calibration ion source. Results of such calibration for the NPA Channel 3 are shown in Fig.7. Curve 1 of this figure was measured when partially damaged foil with through-holes was installed in the NPA. Foil bias voltage in this run was equal to zero, that is the energy of protons on the entrance of the foil is equal to ion accelerating voltage. Two peaks clearly observed on the curve correspond to ions passed through the holes in the foil (without energy loss) and to ions transmitted through the foil and lost part of their initial energy. Observed energy loss about 1.6 keV significantly exceed expected from stopping power value for 14 keV protons (0.46 keV / (g/cm$^2$)) [10] and nominal foil linear density 2 μg/cm$^2$ and correspond to effective foil thickness 17 nm.

There is a probability that foil damage causes the increase of thickness of unbroken parts of the foil. To verify this statement, we performed additional measurements with fresh undamaged foil (curve 2 in fig.7). Foil bias voltage in this run was +8 kV, which means that for ion source accelerating voltage 14 kV the energy of protons on the entrance of the foil is 6 keV. Observed energy loss in the foil for this run is 1.1 kV, stopping power for 6 keV protons is 0.33 keV / (g/cm$^2$), which gives the same effective thickness of 17 nm.

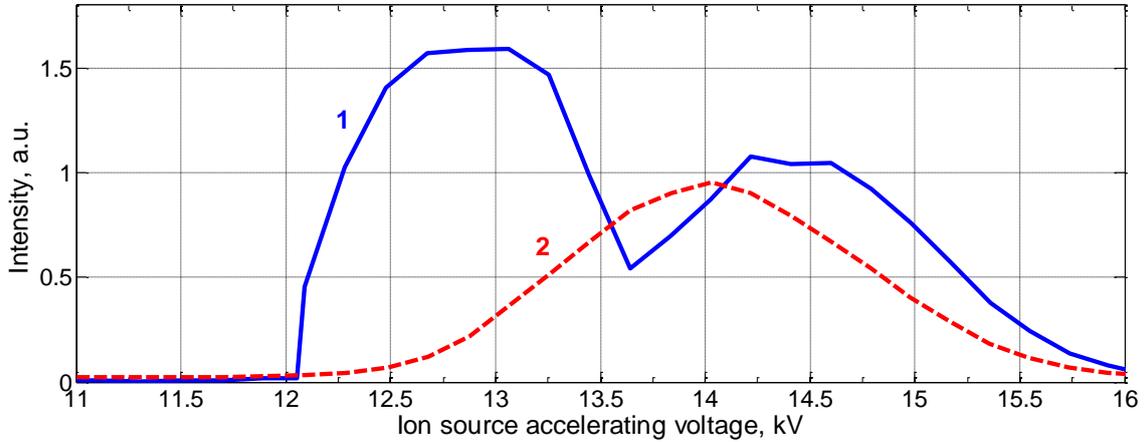

**Figure 8.** NPA channel 3 response for calibration source accelerating voltage scan. **1** – partially damaged foil (with holes), H$^+$ ions, zero foil bias voltage; **2** – undamaged foil, H$^+$ ions, foil bias voltage +8 kV.

NIST database contains the average value of stopping power. NPA predominantly transmits and counts particles with low scattering angles, and this effect can potentially influence observed energy loss in the foil. For investigation of the influence of this effect, we produced a calculation of particle trajectories for different foil thickness for conditions of NPA calibration with partially damaged foil discussed afore (H$^+$ ions, initial ion energy 14.5 keV, zero foil bias voltage). The dependence of particles' spot position from the foil thickness is shown on the fig. 9a. A thickness of 18 nm corresponds to the best hit into the aperture of NPA channel 3, and we will use this value below as an "effective" foil thickness. Figure 9b shows detector surface hit points for two sorts of particles, corresponding two peaks on the curve 1 of fig.8, namely protons with initial energy 12.9 keV passed without interaction with the foil (blue squares) and protons with initial energy 14.5 keV passed through 18 nm stripping foil (red circles).

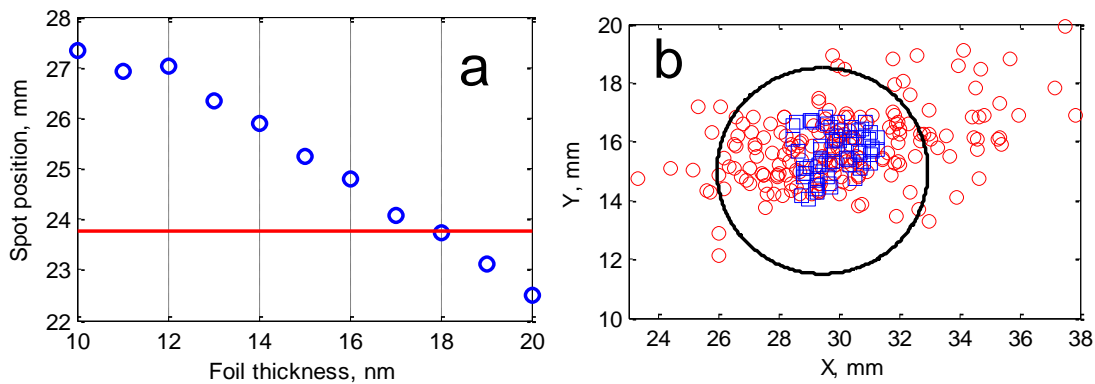

**Figure 9.** NPA Calculation of particle trajectories for different foil thicknesses. **a** – dependence of particle spot position from foil thickness (H$^+$, initial energy 14.5 keV, zero foil bias voltage), horizontal line marks experimentally observed position; **b** – particles hit points on the detector surface, blue squares – without foil, initial H$^+$ ion energy 12.9 keV, red circles – foil thickness 18 nm, initial ion energy 14.5 keV, solid black circle – detector aperture

## 3.3 Particles trajectories calculations

The calculated probability for a particle with certain initial energy to transmit through 18 nm thick carbon foil is shown in fig.10a. This probability as well probabilities of stripping and counting of ions by a detector discussed later is used on the third (postprocessing) stage of calculations for estimation of overall probability to detect an input particle.

Notice that the SRIM code can't predict the charge stage of transmitted particles, so stripping probability is estimated using literature data. It is considered, that equilibrium charge distribution achieves, that is this probability determines by target material and particle velocity on the exit of the target, and doesn't depend on the initial charge of the particle and target thickness. The conversion efficiency for hydrogen and deuterium neutrals based on published data [11],[12] is shown in fig.10b. Further, the probability to count ions by detector take equal to 0.5 for all energies and mass of the ions [13],[14].

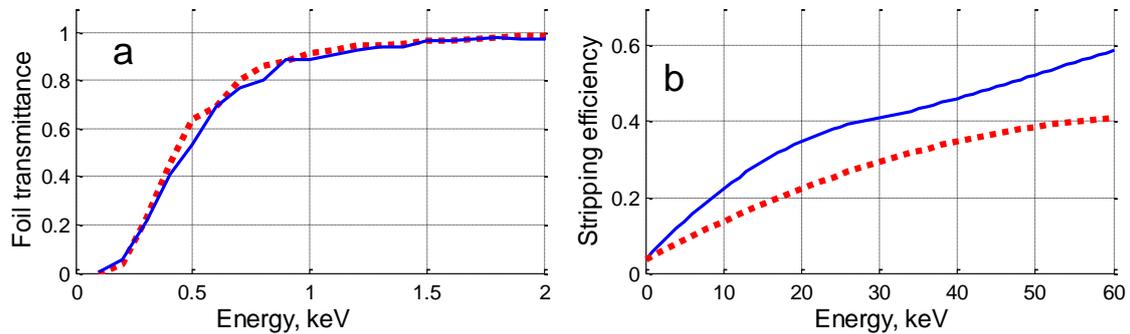

**Figure 10.** a – transmittance of ions through 18 nm carbon foil (SRIM calculations), b – fraction of positive ions ($H^+$, $D^+$) on the exit of the foil (interpolation of published data [11],[12]). Blue solid line – hydrogen, red dot line - deuterium

COMSOL Multiphysics® software [15] is used in the second stage for the calculation of particles trajectories. The start point of the calculated trajectories are distributed over the surface of stripping foil, energies, and angles for each particle are taken from SRIM output.

Totally we have produced calculations of trajectories of 200 particles for each energy up to 50 keV for hydrogen and up to 25 keV for deuterium with energy step 0.4 keV. Instrumental functions on NPA channels are determined as a relative number of trajectories hit the certain detector aperture multiplied to probability to transmit via a stripping foil and to probability to strip (exit the foil in the form of positive ion).

The passing of ions through the foil cause particle angular spread especially important for low-energy particles. In the NPA, particles with big axis angles are cut out using diaphragms on the input and output of a magnet. fractions of particles passed through NPA and achieved detector surface as well lost on the input, output diaphragm and deflector plate, are shown in fig.11.

– 10 –

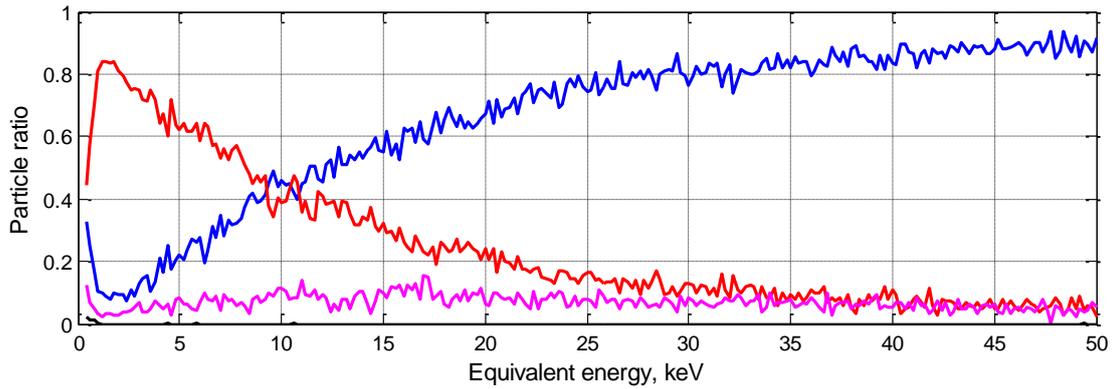

**Figure 11.** Irising of particles in NPA, blue – a fraction of particles achieving detector surface, red –a fraction of particles hit magnet input diaphragm, magenta - a fraction of particles hit magnet output diaphragm, black- a fraction of particles hit deflector plates

The calculated instrumental functions of the channels 1-15 together with measured ones for low-energy channels are shown in fig.12 for hydrogen neutrals ($H^0$) and in fig.13 – for deuterium neutrals ($D^0$). Absolute values for experimental instrumental functions matched for best coincidence with calculations. Energies of the channels, peak efficiencies, and FWHM widths were presented in table 1. The observed differences between calculated and measured values can be explained by the individual sensitivities of the channels. Based on the comparison of calculations with calibration we estimated RMS errors of 35% for channel efficiency and 70 eV for the position of the channels.

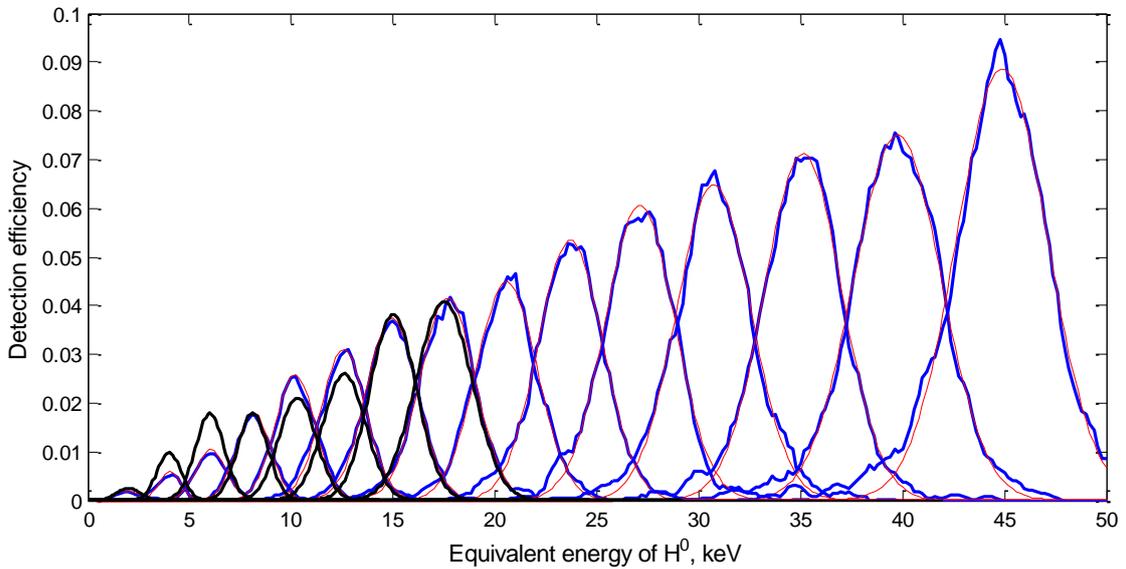

**Figure 12.** Instrumental functions of NPA for $H^0$ particles, stripping foil voltage +8 kV, deflection voltage -8 kV. Blue – calculations, red – gauss fit, black – measurements (data from calibration discussed in section 2.1)

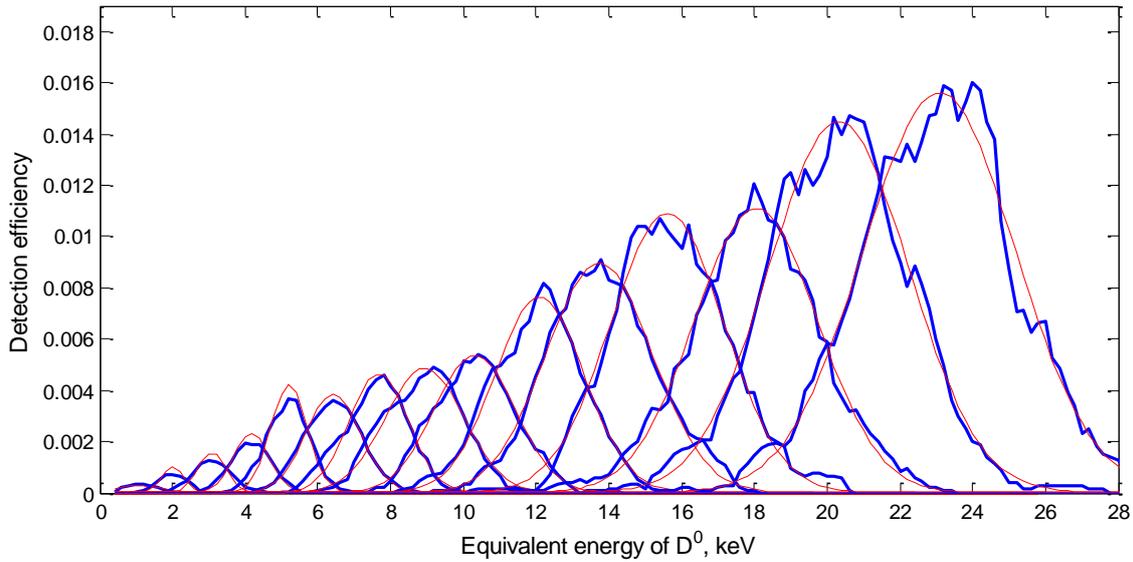

**Figure 13.** Instrumental functions of NPA for $D^0$ particles, stripping foil voltage +4 kV, deflection voltage -4 kV. Blue – calculations, red – gauss fit

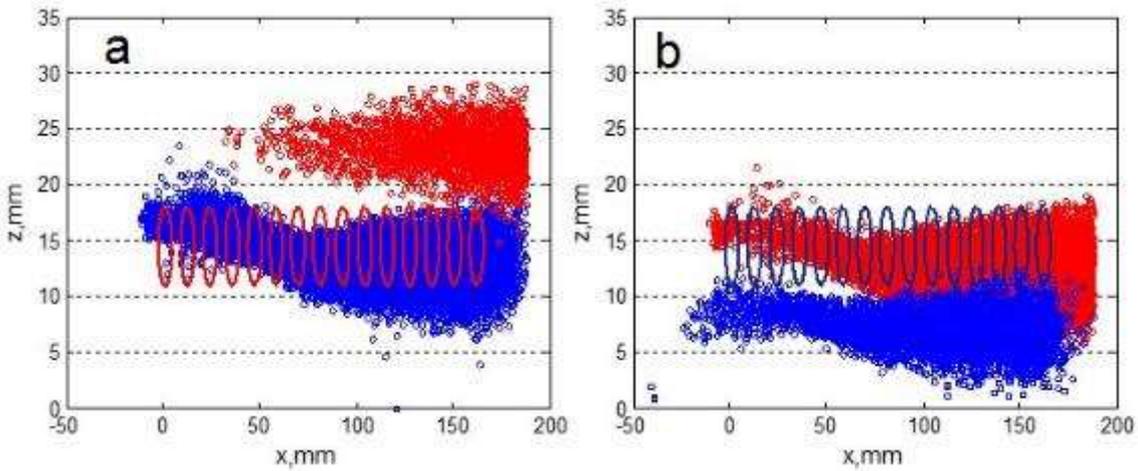

**Figure 14.** Spot diagram of crossing points of particles and detector surface; x-axis – distance along detector surface, y-axis – vertical coordinate of the crossing point, blue points – $H^0$ (energy range 1-50 keV), red points – $D^0$ (energy range 1-30 keV), rings – detector's apertures; **a** - registration of H ions (stripping foil voltage +8 kV, deflection voltage -8 kV), **b** - registration of D ions (stripping foil voltage +4 kV, deflection voltage -4 kV)

### 3.4 Mass separation in the NPA

A feature of the NPA is the possibility to separate particles by their mass and distinguish H or D ions. This option is achieved by a shift of ion trajectories in the NPA in the electrostatic field produced by the deflector plates. For investigation of the capability of NPA to distinguish certain hydrogen isotopes, we produced calculations of particles trajectories for both hydrogen and deuterium for the two modes operation of NPA – detection of H ions (stripping foil voltage +8 kV, deflection voltage -8 kV) and detection of D ions (stripping foil voltage +4 kV, deflection voltage -4 kV). The results of these calculations were shown in fig.14. Spots on the figure mark positions of hitting of detector surface by test particles (blue – H, red - D) with different energies.

As seen from the figure, deflection effectively separates particles, and only a few sporadic particles of rejected species achieve detector aperture.

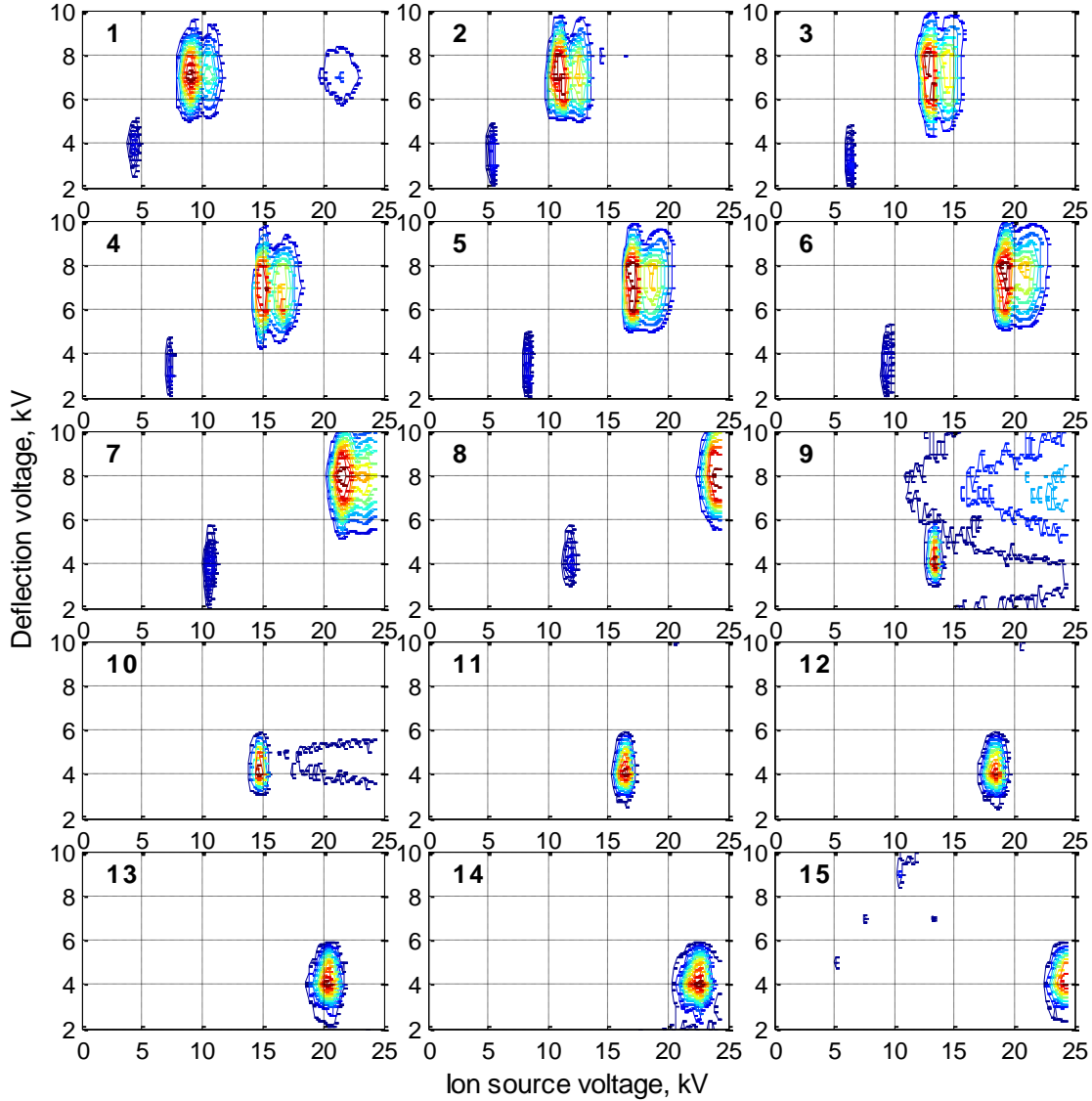

**Figure 15.** Calibration scan over ion source and deflection voltages with partially damaged stripping foil and $U_{foil}=0$. Every plot corresponds to the response of a certain NPA channel (marked by the number at the upper left corner of the plot). Signals from particles with mass m=1 are observed at deflection voltage near 7 kV, signals from particles with mass m=2 - at deflection voltage near 4 kV.

The efficiency of mass separation by the NPA was studied in the calibration runs with partially damaged stripping foil. We produced several scans over ion source accelerating voltage with different voltage on the deflector plates in the range of 2-10 kV and zero stripping foil voltage.

Responses of the NPA channels for different ion source and deflection voltages are presented in Fig.15. The ion source produces $H^+$ and $H_2^+$ ions that are accelerated by a high voltage applied to the ion source. Four particle species are detected by the NPA channels in this operation mode. There are:

- $H^+$ ions passed through holes in the foil without loss of energy. The energy of these ions is equal to the accelerating voltage of the ion source
- $H^+$ ions passed through the foil and lost part of their initial energy (about 2 keV)
- $H^+$ ions appeared from the dissociation of $H_2^+$ ions in the stripping foil
- $H_2^+$ ions passed through holes in the stripping foil

For channel 1 the first specie is observed at ion source voltage near 9 kV and deflection voltage 7 kV, second specie – at voltages 10.5 / 8 kV, third - at voltages 21 / 7 kV, and four – at voltages 4.5 / 4 kV. As seen from fig.15, deflection voltage 4 kV effectively cut off ions with m=1, leaving only ions with m=2 ($H_2^+$) on the signals. In the same way, deflection voltage 8 kV rejects ions with m=2 and leaves only ions with m=1.

## 4. Ion temperature evaluation

The NPA is intended for measurements of the energy distribution of fast hydrogen (deuterium) atoms originated from plasma ions in the reaction of charge exchange in collisions with slow or fast hydrogen atoms. Plasma ion temperature and distribution of fast superthermal ions in plasma can be derived from these measurements.

The neutrals detected by the NPA appear in the charge-exchange collisions

$H^+ + H^0 \rightarrow H^0 + H^+$

There are two possible sources of parent hydrogen atoms in plasma. That is:

a. "Passive target": slow hydrogen atoms originated from fueling gas puffing, proton recombination on the plasma-facing components, or desorption from walls. The energy of these atoms, which is determining by the process of $H_2$ molecule dissociation, is in order of several eV. After getting into plasma the atoms are ionizing by electrons in process $H^0 + e \rightarrow H^+ + e + e$. The reaction rate for this process is about 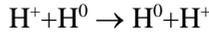$\approx 2 \times 10^{-8}$ cm$^{-3}$s$^{-1}$ for electron temperatures in the range 25 – 2500 eV, which gives characteristic penetration depth of hydrogen atoms in plasma ~3 cm for plasma density $10^{14}$ cm$^{-3}$.

b. "Active target": fast hydrogen atoms are injecting by heating or diagnostic beam injectors (if the pathline of injected neutral beam crosses the volume of plasma observed by NPA).

A rate of production of neutrals matched to acceptance of the NPA is:

$$\frac{dn_{a'}}{dt \cdot dE_{a'}} = \frac{dn_i}{dE_i \cdot d\Omega} n_a \sigma_{cx}(v_{rel}) \cdot v_{rel} \cdot \Delta\Omega$$

$$\frac{dn_i}{dE_i \cdot d\Omega} = f_i \cdot n_i$$

where E – the energy of a primary ion in plasma, $n_a$, $n_i$, $n_{a'}$ – concentrations of atoms, ions, and secondary neutrals, $\sigma_{cx}$ – charge exchange cross-section [16],[17], $v_{rel} = |\vec{v_i} - \vec{v_a}|$ – magnitude of relative velocity of the colliding particles (ion and atom), $\Delta\Omega$ – solid angle of registration, $f_i$ – distribution function of ions in plasma. Here we assumed that momentum transfer does not accompany the charge-exchange process, that is energy and direction on neutrals appearing in the charge-exchange collision is the same as of parent ion. Definitely, the scattering angle for such collisions is not exceeding several degrees (see, for example, [18],[19]).

For the case of passive target velocity of primary is negligible, and

$$v_{rel} = \sqrt{\frac{2E_i}{m_i}}$$

In the second above-mentioned case (active target)

$$v_{rel} = \left(\frac{2E_i}{m_i} + \frac{2E_b}{m_b} - 4\sqrt{\frac{E_i E_b}{m_i m_b}} \cos\theta\right)^{1/2}$$

Where $E_b$, $m_b$ – the energy of the injected target beam and mass of particles in the beam, $\theta$ - the angle between the direction of the beam and line of sight of the NPA.

Hereof, the response (count rate) of certain NPA channel can be represented as:

$$R_k = \int \left(dV \int \left(dE_i \cdot f_i \cdot n_i n_a \sigma_{cx}(v_{rel}) \cdot v_{rel} \cdot (1 - P_{loss}) \cdot \Delta\Omega \cdot I_k(E_i)\right)\right) \quad (1)$$

where V – plasma volume observed by the NPA, $P_{loss}$ – the probability that neutral would be ionized before passing away from plasma, $I_k$ – instrumental function of the channel k. Hereafter we will omit the term (1-$P_{loss}$) for brevity, but it may be substantial for plasma with density exceed $10^{13}$ cm$^{-3}$.

The best way for evaluation of plasma parameters from the NPA data is a comparison of measured response of the NPA channels with results of simulation of plasma by sophisticated codes, which take into account spatial variation plasma parameters and complex behavior of fast ions in the tokamak. Such comparison is out of the scope of this article. Nevertheless, it is useful also to find and exploit parameters of ion distribution function averaged over observed plasma volume which characterizes plasma performance and can be derived straightforwardly from NPA data. Under the assumption of the constancy of the plasma parameters over the observed plasma volume the count rate of NPA can be represented as

$$R_k = V \cdot \Delta\Omega \, \overline{n_i n_a} \int dE_i \cdot f_i(\boldsymbol{P}, E_i) \cdot \sigma_{cx}(v_{rel}(E_i)) \cdot v_{rel}(E_i) \cdot I_k(E_i) = \eta \cdot Y_k(\boldsymbol{P})$$

Here we designated $\boldsymbol{P}$ – a set of parameters of the ion distribution function, $\eta = \overline{n_i n_a}$ – a coefficient common for all channels, which depends on the concentrations of atoms and ions. The functions $Y_k(\boldsymbol{P})$ we will call as response functions of NPA channels. Parameters $\boldsymbol{P}$ as well mean density product $\overline{n_i n_a}$ can be found by minimization of differences between measured and calculated count rates:

$$M = \sum_k w_k^2 \cdot (R_k^{meas} - \eta \cdot Y_k(\boldsymbol{P}))^2 \to \min \quad (2)$$

where $w_k$ – the weight of the certain channel of registration that may be used for taking into account confidence of measurement by the channel.

In the case of the study of plasma without suprathermal ions, an intrinsic parameter of the ion distribution function is an ion temperature $T_i$. In the thermal (Maxwellian) plasma an ion distribution function can be represented as:

$$f_M(T_i, E_i) = \frac{1}{2}\left(\frac{1}{\pi k T_i}\right)^{3/2} exp\left(-\frac{E_i}{k T_i}\right)\sqrt{E_i}$$

Calculated response functions of the NPA channels for Maxwellian plasma and passive target

$$Y_k(\overline{T_i}) = V \cdot \Delta\Omega \cdot \eta \cdot \left(\frac{1}{\pi k \overline{T_i}}\right)^{3/2} \frac{1}{\sqrt{2m_i}} \int dE_i \cdot E_i \cdot exp\left(-\frac{E_i}{k \overline{T_i}}\right) \sigma_{cx}(E_i) \cdot I_k(E_i)$$

are shown in Fig.16. We used here averaging bar over temperature $T_i$ to emphasize that the resulting temperature depends on the distribution of plasma parameters over the observed volume.

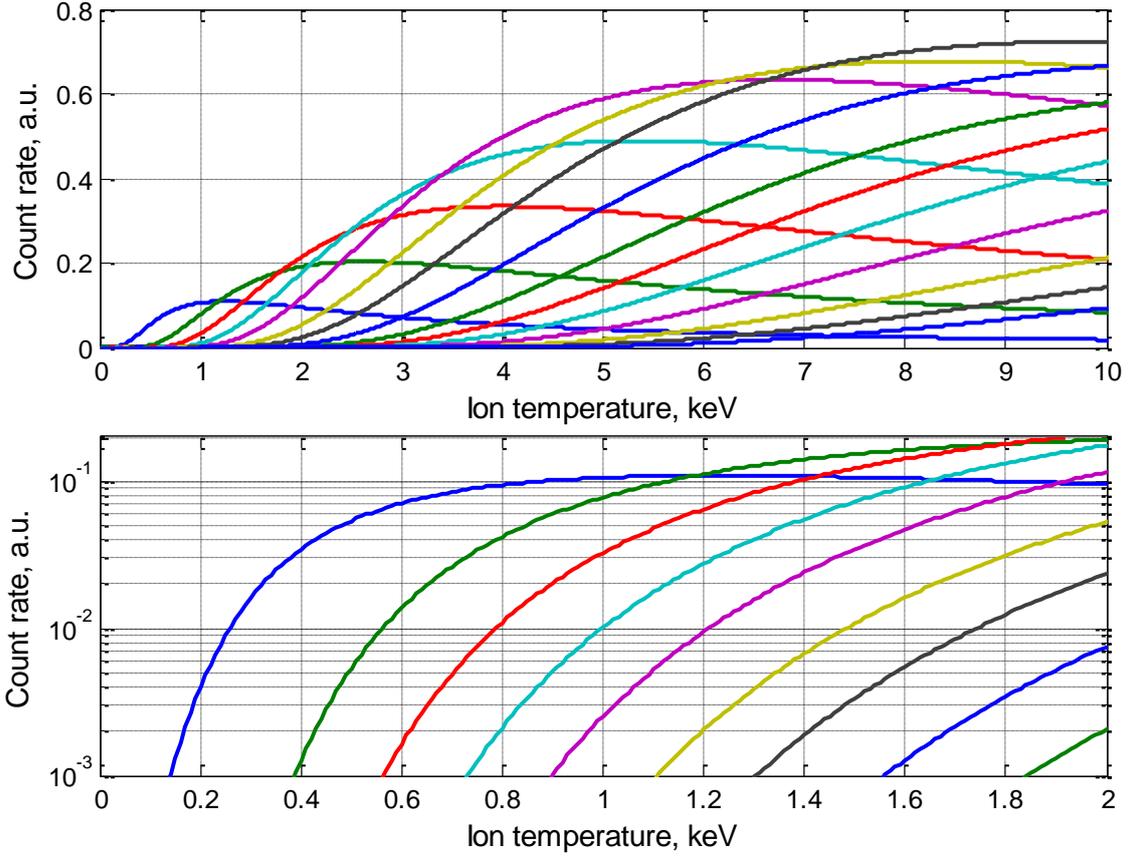

**Figure 16.** Response functions of the NPA channels. Maxwellian hydrogen plasma, passive target, $U_{foil}$=+8 kV. The same response functions are plotted on the top and bottom figures with different axis scales and limits for illustration purposes

Evaluation of ion temperature requires solving of the system of equations:

$$\begin{cases} \dfrac{\partial}{\partial \overline{T_\iota}}\left(\sum_k w_k^2 \cdot (R_k^{meas} - \eta \cdot Y_k(\overline{T_\iota}))^2\right) = 0 \\ \dfrac{\partial}{\partial \eta}\left(\sum_k w_k^2 \cdot (R_k^{meas} - \eta \cdot Y_k(\overline{T_\iota}))^2\right) = 0 \end{cases}$$

or

$$\begin{cases} \eta \sum_k w_k^2 \cdot Y_k \dfrac{\partial Y_k}{\partial \overline{T_\iota}} = \sum_k w_k^2 \cdot R_k^{meas} \cdot \dfrac{\partial Y_k}{\partial \overline{T_\iota}} \\ \eta \sum_k w_k^2 \cdot (Y_k)^2 = \sum_k w_k^2 \cdot R_k^{meas} \cdot Y_k \end{cases} \quad (3)$$

After extraction of the term $\eta$ we receive the equation

$$dM = \dfrac{\sum_k w_k^2 \cdot Y_k \dfrac{\partial Y_k}{\partial \overline{T_\iota}}}{\sum_k w_k^2 \cdot (Y_k)^2} \cdot \sum_k w_k^2 \cdot R_k^{meas} \cdot Y_k - \sum_k w_k^2 \cdot R_k^{meas} \cdot \dfrac{\partial Y_k}{\partial \overline{T_\iota}} = 0 \quad (4)$$

A root of this equation is the desired ion temperature $T_i^{des}$. Second unknown variable $\eta^{des}$ - can be found by substitution of derived value of ion temperature to the system (3).

After determination of values of desirable parameters $T_i$ and $\eta^{des}$, we can estimate the tolerance of this evaluation from the predicted uncertainties of the NPA channels responses

$$(\Delta T_i)^2 = \left(\frac{\partial(dM)}{\partial T_i}\right)^{-2} \sum_k (\Delta R_k^{meas})^2 \left(\frac{\partial(dM)}{\partial R_k^{meas}}\right)^2 \quad (5)$$

The fundamental source of the uncertainty $\Delta R_k^{meas}$ is a random origin of release of neutrals from plasma and and a finite number of counted ions. According central-limit theorem, "statistical" uncertainty is equal to the square root of a number of counted particles

$$\Delta R_k^{meas} = (R_k^{meas})^{0.5}$$

Another way for estimation of measurement tolerance is the use of deviation of measured and expected values of NPA channels responses, which characterized by the quadratic form M

$$(\Delta T_i)^2 = M(T_i^{des}, \eta^{des}) \cdot \left(\frac{\partial(dM)}{\partial T_i}\right)^{-1} \quad T_i = T_i^{des} \quad (6)$$

This estimation includes both statistical and systematic errors of the measurements.

Let's notice, that formally the described procedure allow to found two parameters of plasma - mean ion temperature $\overline{T_i}$ and mean product of concentrations of atoms and ions in plasma $\overline{n_i n_a}$. At the same time, we did not produced absolute calibration of the NPA, and our simulation of NPA response contains several insufficiently accurate parameters taken from literature data (such as probability to count ions by detector), therefore only dynamics, but not an absolute value of the parameter $\overline{n_i n_a}$ can be derived now from the NPA measurements.

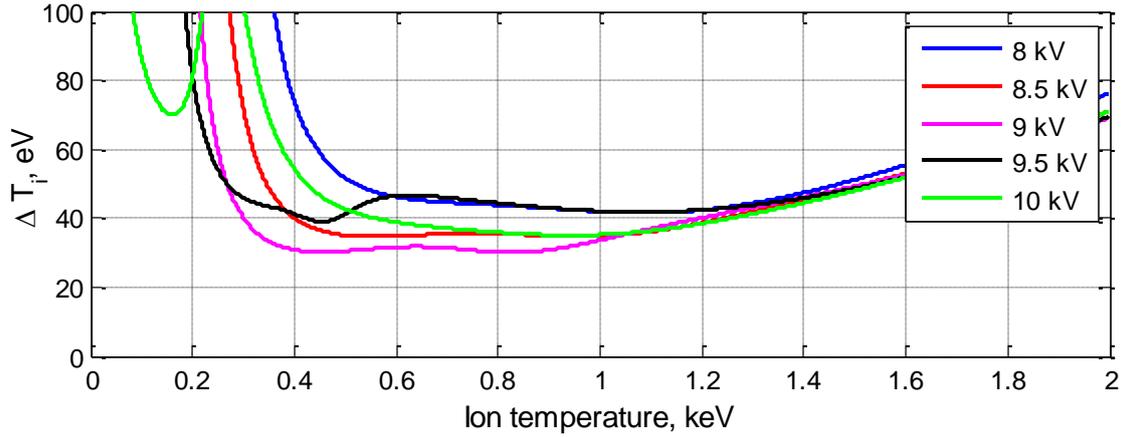

**Figure 17.** Estimation of ion temperature measurement uncertainty for different stripping foil voltages. See detailed description in the text

A rather specialised task for the NPA is a measurement of low ion temperature (below 1 keV). In this case signals are present in only a few low-energy channels, and postions of these channels on the energy scale can strongly influence to an accuracy of the measurements. The energy scale of the channels can be shifted by changing of the stripping foil voltage which accelerates the ions in the NPA. The magnitude of this voltage is subject to optimization.

We have produced calculations of instrumental functions of the NPA channels and response functions for Maxwellian plasma according to the above procedure for five values of stripping foil voltage (+8, +8.5, +9, +9.5, and +10 kV) and have estimated ion temperature unsertanties for specified stripping foil voltages and ion temperature range 0-2 keV (Fig. 17). Predicted ion temperature uncertainty is estimaped according the formulas (2) - (5) in asumption of statistical errors of the measurements and discrete nature of counting. Based on this assumption, we adopted the weights of the registration channels is proportional to square root of the channel signal

$$w_k^2 = R_k^{meas}$$

and use the follows estimation of uncertainties of measurements of the signals:

$$\Delta R_k^{meas} = (R_k^{meas})^{0.5} + 1$$

In the such model the measurement uncertainty depends on a signal level (number of counts) of the NPA channels, which in turn depends on the geometry of the system and mean density product $\overline{n_i n_a}$. The unsertainty on Fig.17 was calculated for actual geometry of the NPA installed on ST40, mean density product $\overline{n_i n_a} = 10^{20} cm^{-6}$ and measuremet duration 1 ms. As seen in Figure 17, optimal stripping foil voltage for low-temperature measurements is +9 kV. Let's notice additionally, that while the level of unsertainty shown in Fig.17 depends on the signal values, the shapes of the plotted curves and thus optimal voltage only weakly depend on signal level and the used model of unsertainty.

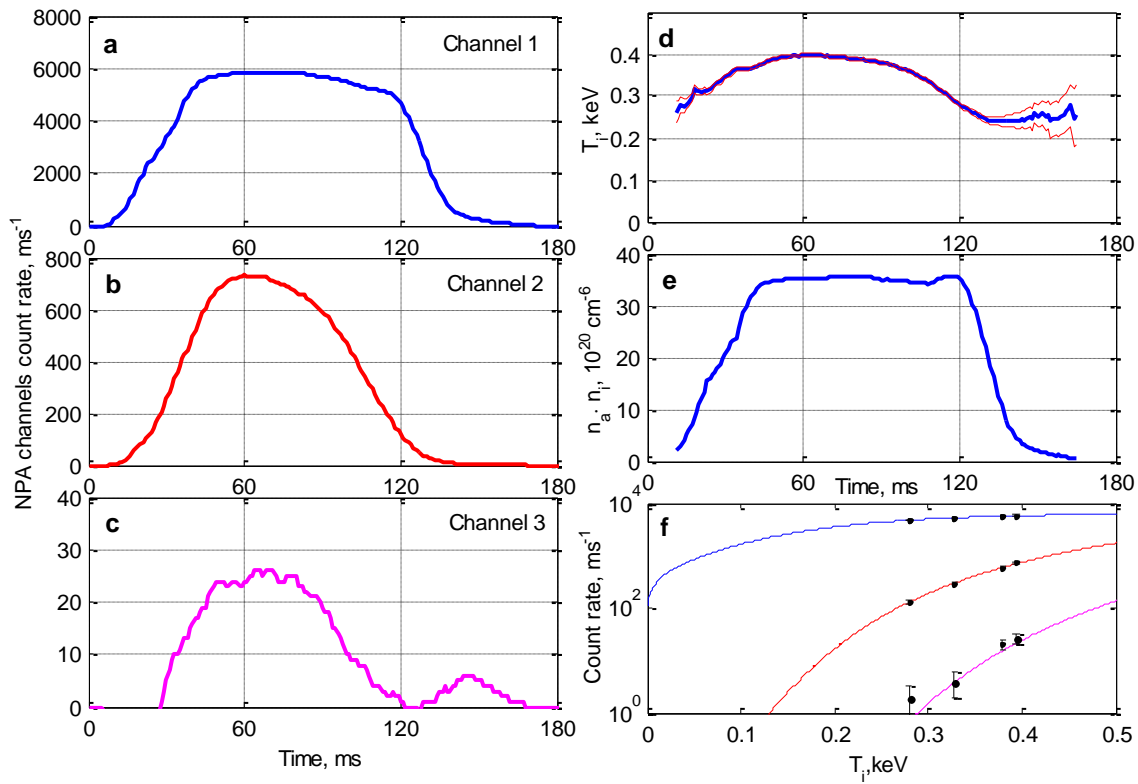

**Figure 18.** ST40 shot #6707: Measurements of ion temperature in low-temperature plasma. **a,b,c** – count rates for first three channels of the NPA; ,**d,e** – derived ion temperature and mean product of density of ions to density of neutrals, thin red lines represent the band of statistical uncertanty; **f** – instrumental functions of the NPA channels (color solid lines) with the values of measured count rates (black dots with error bars)

An example of use of the NPA for measurements in low-temperature plasma is shown in Fig.18 (ST40 shot #6707). Stripping foil voltage was +9 kV in this measurement. Useful signals are observed in the first three channels of NPA. Plots 18 **a-c** represents count rates of the NPA channels after some conditioning and background elimination. Plots 18 **d** and **e** demonstrate derived ion temperature and the neutral to ion density product $\overline{n_i n_a}$. Thin red lines in the Fig.18d demonstrates the band of statistical uncertanty. Let's notice again that due lack of absolute sensitivity calibration and presence of several obscure calibration factors results of density measurements (Fig.18e) can be used only for study of relative dynamics of density.

The response functions of NPA (color lines) together with measured count rates (black dots with error bars) for several moments with different ion temperature are shown in Fig.18f. We do not observe any discrepancies outside of statistical errors which indicates the absence of significant systematic errors.

## Conclusion

The neutral particle analyzer designed and installed on ST40 tokamak enables the measurement of energy distribution functions for hydrogen ions (up to 40 keV) and deuterium ions (up to 20 keV), as well as the ion temperature of the plasma at temperatures above 300 eV.

## Acknowledgments

The authors thank A.A.Ivanov for initiating the work on the neutral analyzer for the ST40 tokamak and M.Gryaznevich for his interest in the work and support in preparing the article.

The authors would like to deeply thank V. Nemytov, C. Bradley, and the rest of the Tokamak Energy team for their assistance in NPA calibration and experiments on the ST40 tokamak.

## Declaration of competing interest
The authors declare that they have no known competing financial interests or personal relationships that could have appeared to influence the work reported in this paper.

## Data availability
The data presented in the figures of this article are available in the form of MATLAB figures at Mendeley Data service DOI: 10.17632/ht3htpcz74.1.

Authors will be happy to provide data of interest in other formats upon request.

## Funding sources

This research did not receive any specific grant from funding agencies in the public, commercial, or not-for-profit sectors.

## CRediT authorship contribution statement

**S. Polosatkin:** Writing – original draft, Writing – review & editing, Visualization, Investigation, Validation, Supervision, Software, Methodology, Formal analysis.
**A. Belykh:** Writing – review &editing, Investigation
**A. Rovenskikh**: Writing – review & editing, Investigation, Software.